\documentstyle[aps,prl,psfig]{revtex}
\makeatletter
\def\setcaption#1{\def\@captype{#1}}
\makeatother

\newcommand{\degree}    {^{\circ}}
\newcommand{\nue}       {{\nu}_{\rm e}}
\newcommand{\nuebar}    {\bar{\nu}_{\rm e}}
\newcommand{\numu}      {\nu_{\rm \mu}}
\newcommand{\numubar}   {\bar\nu_{\rm \mu}}
\newcommand{\nutau}     {{\nu}_{\rm \tau}}

\newcommand{\sstt}      {\sin^2 2\theta}
\newcommand{\dms}       {\Delta m^2}
\newcommand{\Frejus}    {Fr\'{e}jus}
\newcommand{\SuperK}    {Super--Kamiokande}
\newcommand{\gsim}      {\mathrel{\rlap{\raisebox{0.3ex}{$>$}}\raisebox{-0.6ex}{$\sim$}}}

\begin{document}
\bigskip
\medskip
{\center \Large
Evidence for oscillation of atmospheric neutrinos\\
}
\bigskip
{\center \large The Super-Kamiokande Collaboration\\}

%
%

\begin{center}
\newcounter{foots}
Y.Fukuda$^a$, T.Hayakawa$^a$, E.Ichihara$^a$, K.Inoue$^a$,
K.Ishihara$^a$, H.Ishino$^a$, Y.Itow$^a$,
T.Kajita$^a$, J.Kameda$^a$, S.Kasuga$^a$
% ...

\footnotesize \it

$^a$Institute for Cosmic Ray Research, University of Tokyo, Tanashi,
Tokyo 188-8502, Japan\\
% ...
\end{center}

\begin{abstract} We present an analysis of atmospheric neutrino data
from a 33.0 kiloton-year (535-day) exposure of the \SuperK{}
detector. The data exhibit a zenith angle dependent deficit of muon
neutrinos which is inconsistent with expectations based on
calculations of the atmospheric neutrino flux. Experimental biases and
uncertainties in the prediction of neutrino fluxes and cross sections
are unable to explain our observation. The data are consistent,
however, with two-flavor $\numu \leftrightarrow \nutau$ oscillations
with $\sstt>0.82$ and $5 \times 10^{-4} < \dms < 6 \times 10^{-3}$
eV$^2$ at 90\% confidence level.
\end{abstract}

\twocolumn
Atmospheric neutrinos are produced as decay products in hadronic
showers resulting from collisions of cosmic rays with nuclei in the upper
atmosphere. Production of electron and muon neutrinos is dominated by
the processes $\pi^+ \rightarrow \mu^+ + \numu$ followed by $\mu^+
\rightarrow e^+ + \numubar + \nue$ (and their charge conjugates) giving
an expected ratio ($\equiv \numu/\nue$) of the flux of $\numu +
\numubar$ to the flux of $\nue + \nuebar$ of about two. The $\numu/\nue$ 
ratio has been calculated in detail with an uncertainty
of less than 5\% over a broad range of energies from 0.1 GeV to 10
GeV~\cite{gaisserflx,hondaflx}.

The $\numu/\nue$ flux ratio is measured in deep underground
experiments by observing final-state leptons produced via
charged-current interactions of neutrinos on nuclei, $\nu + N
\rightarrow l + X$. The flavor of the final state lepton is used to
identify the flavor of the incoming neutrino.

The measurements are reported as $R \equiv
(\mu/e)_{DATA}/(\mu/e)_{MC}$, where $\mu$ and $e$ are the number of
muon-like ($\mu$-like) and electron-like ($e$-like) events
observed in the detector for both
data and Monte Carlo simulation. This ratio largely cancels
experimental and theoretical uncertainties, especially the uncertainty
in the absolute flux. $R=1$ is expected if the physics in the Monte
Carlo simulation accurately models the data.
Measurements of significantly small values
of $R${} have been reported by the deep underground water
Cherenkov detectors Kamiokande\cite{kam_atmanomaly,kam_zenith},
IMB\cite{imb_atmanomaly}, and recently by
\SuperK~\cite{sk_subgev,sk_multigev}. Although measurements of $R${}
by early iron-calorimeter experiments
\Frejus\cite{frejus} and NUSEX\cite{nusex} with smaller data samples
were consistent with 
expectations, the Soudan-2 iron-calorimeter experiment has reported
observation of a small value of $R$\cite{soudan_atmanomaly}.

Neutrino oscillations have been suggested to explain measurements of
small values of $R$. For a two-neutrino oscillation hypothesis, the
probability for a neutrino produced in flavor state $a$ to be observed
in flavor state $b$ after traveling a distance $L$ through a vacuum
is:
\begin{equation}
P_{a \rightarrow b} = \sstt \sin^2{\Big(}\frac{1.27 
\dms(\textrm{eV}^2) L (\textrm{km})}{E_\nu(\textrm{GeV})}\Big{)},
\end{equation}
where $E_\nu$ is the neutrino energy, $\theta$ is the mixing angle
between the flavor eigenstates and the mass eigenstates, and $\dms$ is
the mass-squared difference of the neutrino mass eigenstates. For
detectors near the surface of the Earth, the neutrino flight distance,
and thus the oscillation probability, is a function of the zenith angle of
the neutrino direction. Vertically downward-going neutrinos travel
about 15 km while vertically upward-going neutrinos travel about
13,000 km before interacting in the detector. The broad energy spectrum 
and this range of neutrino flight distances
makes measurements of atmospheric neutrinos sensitive to neutrino
oscillations with $\dms$ down to $10^{-4}$ eV$^2$. The zenith angle
dependence of $R${} measured by the Kamiokande experiment at high
energies has been cited as evidence for neutrino
oscillations\cite{kam_zenith}.

We present our analysis of 33.0 kiloton-years (535 days) of
atmospheric neutrino data from \SuperK. In addition to measurements of
small values of $R${} both above and below $\sim$1 GeV, we observed
a significant zenith angle dependent deficit of $\mu$-like
events. While no combination of known uncertainties in the
experimental measurement or predictions of atmospheric neutrino fluxes
is able to explain our data, a two-neutrino oscillation model of
$\numu \leftrightarrow \nu_x$, where $\nu_x$ may be $\nutau$ or a new,
non-interacting ``sterile'' neutrino, is consistent with the observed 
flavor ratios and zenith angle distributions over the entire energy region. 

\SuperK{} is a 50 kiloton water Cherenkov detector instrumented with
11,146 photomultiplier tubes (PMTs) facing an inner 22.5 kiloton
fiducial volume of ultra-pure water. Interaction kinematics are
reconstructed using the time and charge of each PMT signal. The inner
volume is surrounded by a $\sim$2 meter thick outer detector
instrumented with 1885 outward-facing PMTs. The outer detector is used
to veto entering particles and to tag exiting tracks.

\SuperK{} has collected a total of 4353 fully-contained (FC)
events and 301 partially-contained (PC) events in a 33.0 kiloton-year
exposure. FC events deposit all of their Cherenkov light in the inner
detector while PC events have exiting tracks which deposit some
Cherenkov light in the outer detector. For this analysis, the neutrino
interaction vertex was required to have been reconstructed within the
22.5 kiloton fiducial volume, defined to be $>2$~m from the PMT wall.

FC events were separated into those with a single visible Cherenkov
ring and those with multiple Cherenkov rings.  For the analysis of FC
events, only single-ring events were used.  Single-ring events were
identified as $e$-like or $\mu$-like based on a likelihood analysis of
light detected around the Cherenkov cone.  The FC events were
separated into ``sub-GeV'' ($E_{vis}<1330$ MeV) and ``multi-GeV''
($E_{vis}>1330$ MeV) samples, where $E_{vis}$ is defined to be the
energy of an electron that would produce the observed amount of
Cherenkov light. $E_{vis}=1330{}$ MeV corresponds to $p_\mu \sim
1400{}$ MeV$/c$.

In a full-detector Monte Carlo simulation, 88\% (96\%) of the sub-GeV
$e$-like ($\mu$-like) events were $\nue$ ($\numu$) charged-current
interactions and 84\% (99\%) of the multi-GeV $e$-like ($\mu$-like)
events were $\nue$ ($\numu$) charged-current interactions. PC events
were estimated to be 98\% $\numu$ charged-current interactions; hence,
all PC events were classified as $\mu$-like, and no single-ring
requirement was made. Table~\ref{tb:evts} summarizes the number of
observed events for both data and Monte Carlo as well as the $R$
values for the sub-GeV and multi-GeV samples. Further details of the 
detector, data selection and event reconstruction used in this analysis are
given elsewhere\cite{sk_subgev,sk_multigev}.

\begin{table}
 \begin{center}
\begin{tabular}{lrrrr} & Data & Monte Carlo & \\
  \hline
  \multicolumn{3}{l}{sub-GeV} & \\ 
  {} single-ring   & 2389 & 2622.6 \ \ \ \ & \\
  {} {} $e$-like   & 1231 & 1049.1 \ \ \ \ & \\ 
  {} {} $\mu$-like & 1158 & 1573.6 \ \ \ \ & \\ 
  {} multi-ring    &  911 &  980.7 \ \ \ \ & \\ 
  \hline 
  total            & 3300 & 3603.3 \ \ \ \ & \\ 
  \multicolumn{1}{r}{$R=$} & 
    \multicolumn{3}{l}{$0.63~\pm~0.03~(stat.)~\pm~0.05~(sys.)$} \\ 
  \hline \hline 
  \multicolumn{3}{l}{multi-GeV} & \\ 
  {} single-ring   & 520  & 531.7 \ \ \ \ & \\ 
  {} {} $e$-like   & 290  & 236.0 \ \ \ \ & \\ 
  {} {} $\mu$-like & 230  & 295.7 \ \ \ \ & \\ 
  {} multi-ring    & 533  & 560.1 \ \ \ \ & \\ 
  \hline 
  total            & 1053 & 1091.8 \ \ \ \ & \\	
  \hline  
  {} partially-contained & 301 & 371.6 \ \ \ \ & \\ 
  \multicolumn{1}{r}{$R_{FC+PC}=$} & 
    \multicolumn{3}{l}{$0.65~\pm~0.05~(stat.)~\pm~0.08{}~(sys.)$} \\
  \end{tabular} 
  \end{center} 
  \caption{Summary of the sub-GeV, multi-GeV and PC event samples
  compared with the Monte Carlo prediction based on the neutrino flux
  calculation of Ref.~[2].}
\label{tb:evts} 
\end{table}

We have measured significantly small values of $R$ in both the sub-GeV
and multi-GeV samples. Several sources of systematic uncertainties in
these measurements have been considered. Cosmic ray induced
interactions in the rock surrounding the detector have been suggested
as a source of $e$-like contamination from neutrons, which could 
produce small $R$ values~\cite{ryazh},
but these backgrounds have been shown to be
insignificant for large water Cherenkov detectors\cite{kam_neutrons}.
In particular, \SuperK{} has 4.7 meters of water surrounding the
fiducial volume; this distance corresponds to roughly 5 hadronic
interaction lengths and 13 radiation lengths. Distributions of event
vertices exhibit no excess of $e$-like events close to the fiducial
boundary\cite{sk_subgev,sk_multigev}.

The prediction of the ratio of the $\numu$ flux to the $\nue$ flux is
dominated by the well-understood decay chain of mesons and contributes
less than 5\% to the uncertainty in $R$. Different neutrino flux
models vary by about $\pm$20\% in the prediction of absolute rates, but the
ratio is robust\cite{flxcompare}. Uncertainties in $R$ due to a
difference in cross sections for $\nue$ and $\numu$ have been
studied\cite{engle}; however, lepton universality prevents any
significant difference in cross-sections at energies much above the
muon mass and thus errors in cross-sections could not produce a small
value of $R$ in the multi-GeV energy range. Particle identification
was estimated to be $\gsim 98\%$ efficient for both $\mu$-like and
$e$-like events based on Monte Carlo studies. Particle identification
was also tested in \SuperK{} on Michel electrons and stopping
cosmic-ray muons and the $\mu$-like and $e$-like events used in this
analysis are clearly separated\cite{sk_subgev}. The particle
identification programs in use have also been tested using beams of
electrons and muons incident on a water Cherenkov detector at
KEK\cite{kam_kek}.
The data
have been analyzed independently by two groups, making the possibility
of significant biases in data selection or event reconstruction
algorithms remote\cite{sk_subgev,sk_multigev}.
Other explanations for the small value
of $R$, such as contributions from nucleon decays~\cite{mann},
can be discounted
as they would not contribute to the zenith angle effects described
below.

We estimate the probability that the observed $\mu/e$ ratios could
be due to statistical fluctuation is less than 0.001\% for sub-GeV $R$
and less than 1\% for multi-GeV $R$.


The $\mu$-like data exhibit a strong asymmetry in zenith angle
($\Theta$) while no significant asymmetry is observed in the $e$-like
data~\cite{sk_multigev}.
The asymmetry is defined as $A = (U-D)/(U+D)$ where $U$ is the
number of upward-going events ($-1 <
\cos \Theta < -0.2$) and $D$ is the number of downward-going events 
($0.2 < \cos \Theta < 1$). The asymmetry is expected to be near zero
independent of flux model for $E_\nu > 1$ GeV, above which effects due
to the Earth's magnetic field on cosmic rays are small. Based on a
comparison of results from our full Monte Carlo simulation using
different flux models\cite{gaisserflx,hondaflx} as inputs, treatment
of geomagnetic effects results in an uncertainty of roughly $\pm 0.02$
in the expected asymmetry of $e$-like and $\mu$-like sub-GeV events
and less than $\pm 0.01$ for multi-GeV events.  Studies of decay
electrons from stopping muons show at most a $\pm 0.6\%$ up-down
difference in Cherenkov light detection\cite{asymimprove}. 

\begin{figure}[t]
\begin{center}
\psfig{figure=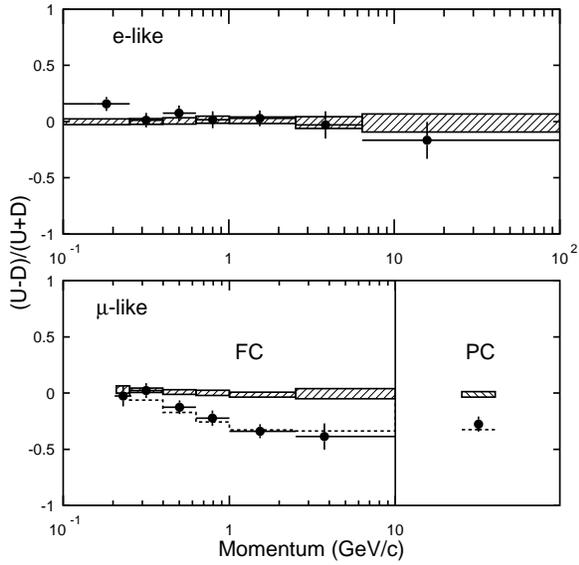,width=3.25in}
\caption{The $(U-D)/(U+D)$ asymmetry as a function of momentum for FC
$e$-like and $\mu$-like events and PC events. While it is not possible
to assign a momentum to a PC event, the PC sample is estimated to have
a mean neutrino energy of 15 GeV. The Monte Carlo expectation
without neutrino oscillations is shown
in the hatched region with statistical and systematic errors added in
quadrature. The dashed line for $\mu$-like is the expectation for $\numu
\leftrightarrow \nutau$ oscillations with $(\sstt=1.0,
\dms = 2.2 \times 10^{-3}$ eV$^2$).}
\label{fig:asym} 
\end{center}
\end{figure}

Figure~\ref{fig:asym} shows $A$ as a function of momentum for both
$e$-like and $\mu$-like events. In the present data, the asymmetry as
a function of momentum for $e$-like events is consistent with
expectations, while the $\mu$-like asymmetry at low momentum is
consistent with zero but significantly deviates from expectation at
higher momentum.
The average angle between
the final state lepton direction and the incoming neutrino direction
is $55\degree$ at $p = 400$ MeV/$c$ and $20\degree$ at 1.5
GeV/$c$. 
At the lower 
momenta in Fig.~\ref{fig:asym}, the possible asymmetry of the neutrino flux 
is largely washed out. 
We have found no detector bias differentiating 
$e$-like and $\mu$-like events that could explain an asymmetry in
$\mu$-like events but not in $e$-like events~\cite{sk_multigev}.

Considering multi-GeV (FC+PC) muons alone, the measured asymmetry, $A
= -0.296 \pm 0.048 \pm 0.01$ deviates from zero by more than 6
standard deviations.

We have examined the hypotheses of two-flavor $\numu \leftrightarrow
\nue$ and $\numu \leftrightarrow \nutau$ oscillation models using a
$\chi^2$ comparison of data and Monte Carlo, allowing all important
Monte Carlo parameters to vary weighted by their expected uncertainties.

The data were binned by particle type, momentum, and $\cos
\Theta$. A $\chi^2$ is defined as:
{\setlength\arraycolsep{2pt}
\begin{eqnarray}
\chi^2 & = & \sum_{cos\Theta, p}
(N_{DATA} - N_{MC})^2/ \sigma^2 + \sum_j \epsilon_j^2/\sigma_j^2,
\end{eqnarray}}
\noindent
where the sum is over five bins equally spaced in $\cos \Theta$
and seven
momentum bins for both $e$-like events and $\mu$-like plus PC events
(70 bins total).
The statistical error, $\sigma$, accounts for
both data statistics and the weighted Monte Carlo statistics.
$N_{DATA}$ is the measured number of events in each
bin. $N_{MC}$ is the weighted sum of Monte Carlo events:
\begin{eqnarray}
N_{MC} & = &
\frac{{\mathcal L}_{DATA}}{{\mathcal L}_{MC}} \times
\sum_{\mbox{MC events}}{\hspace{-0.2in}w}.
\end{eqnarray}
\noindent
${\mathcal L}_{DATA}$ and ${\mathcal L}_{MC}$ are the data and
Monte Carlo live-times. 
For each Monte Carlo event, the weight $w$ is given by:
{\setlength\arraycolsep{2pt} {
\begin{eqnarray} w & = & (1+\alpha)(E^i_\nu/E_0)^\delta 
  (1+\eta_{s,m}\cos\Theta) \times \nonumber \\
&& f_{e,\mu}(\sstt,\dms, (1+\lambda)L/E_\nu) \nonumber \\ 
& \times & { \left\{ \begin{array}{lr} 
(1-\beta_s/2)&\textrm{sub-GeV $e$-like}\\ 
(1+\beta_s/2)&\textrm{sub-GeV $\mu$-like}\\ 
(1-\beta_m/2)&\textrm{multi-GeV $e$-like}\\
(1+\beta_m/2)(1-\frac{\rho}{2}\frac{N_{PC}}{N_{\mu}}) 
  & \textrm{multi-GeV $\mu$-like} \\ 
(1+\beta_m/2)(1+\frac{\rho}{2}). & \textrm{PC} \\
\end{array} \right. } \nonumber \\ 
\end{eqnarray}}}
\noindent
$E_\nu^i$ is the average neutrino energy in the $i^{\rm \small th}$
momentum bin; $E_0$ is an arbitrary reference energy (taken to be 2
GeV); $\eta_{s}$ ($\eta_{m}$) is the up-down uncertainty of
the event rate in the
sub-GeV (multi-GeV) energy range; $N_{PC}$ is the total number of Monte
Carlo PC events; $N_\mu$ is the total number of Monte Carlo FC
multi-GeV muons. The factor $f_{e,\mu}$ weights an event accounting
for the initial neutrino fluxes (in the case of $\numu \leftrightarrow
\nue$), oscillation parameters and $L/E_\nu$. The meaning of the Monte
Carlo fit parameters, $\alpha$ and $\epsilon_j \equiv $($\beta_s$,
$\beta_m$, $\delta$, $\rho$, $\lambda$, $\eta_s$, $\eta_m)$ and their
assigned uncertainties, $\sigma_{j}$,
are summarized in Table~\ref{tb:parms}. The
over-all normalization, $\alpha$, was allowed to vary freely. The
uncertainty in the Monte Carlo $L/E_\nu$ ratio ($\lambda$) was
conservatively estimated based on the uncertainty in absolute energy
scale, uncertainty in neutrino-lepton angular and energy correlations,
and the uncertainty in production height. The oscillation simulations
used profiles of neutrino production heights calculated in
Ref.~\cite{prodhgt}, which account for the competing factors of
production, propagation, and decay of muons and mesons through the
atmosphere. For $\numu
\leftrightarrow \nue$, effects of matter on neutrino propagation
through the Earth were included following
Ref.~\cite{wolfenstein,mikhsmir}. Due to the small number of events
expected from $\tau$-production, the effects of $\tau$ appearance and
decay were neglected in simulations of $\numu \leftrightarrow
\nutau$. A global scan was made on a $(\sstt, \log \dms)$ grid
minimizing $\chi^2$ with respect to $\alpha$, $\beta_s$, $\beta_m$,
$\delta$, $\rho$, $\lambda$, $\eta_s$ and $\eta_m$ at each point.

\begin{figure}[b]
\begin{center}
\psfig{figure=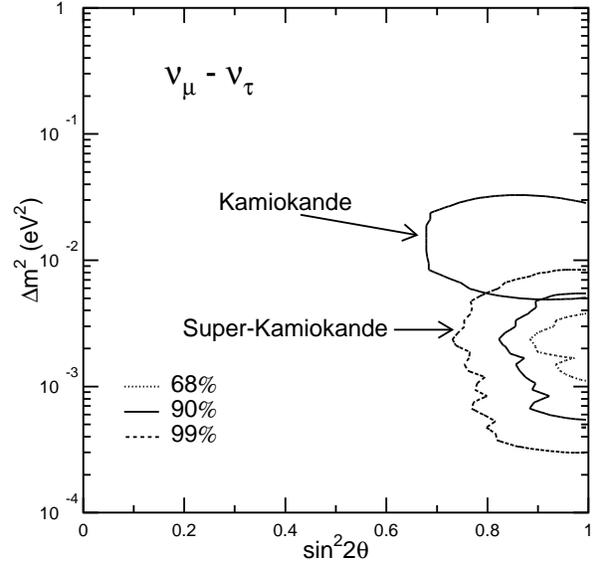,width=3.375in}
\caption{The 68\%, 90\% and 99\% confidence intervals are shown for
$\sstt$ and $\dms$ for $\numu \leftrightarrow \nutau$ two-neutrino
oscillations based on 33.0 kiloton-years of \SuperK{} data. The 90\%
confidence interval obtained by the Kamiokande experiment is also
shown.}
\label{fig:allowed} 
\end{center}
\end{figure}

The best-fit to $\numu \leftrightarrow \nutau$ oscillations,
$\chi^2_{min} = 65.2 / 67 {\rm ~DOF}$,  was
obtained at $(\sstt = 1.0, \dms = 2.2\times 10^{-3} $~eV$^2$) inside
the physical region ($0 \leq \sstt \leq 1$). The best-fit values of
the Monte Carlo parameters (summarized in Table~\ref{tb:parms}) were
all within their expected errors. The global minimum occurred slightly
outside the physical region at $(\sstt = 1.05,
\dms = 2.2\times 10^{-3}~$eV$^2, \chi^2_{min} = 64.8/67~{\rm
DOF})$. The contours of the 68\%, 90\% and 99\% confidence intervals
are located at $\chi^2_{min} + 2.6, 5.0,$ and $9.6$ based on the
minimum inside the physical region\cite{pdgrenorm}. These contours are
shown in Fig.~\ref{fig:allowed}. The region near $\chi^2$ minimum is
rather flat and has many local minima so that inside the 68\% interval
the best-fit $\dms$ is not well constrained. Outside the 99\% allowed
region the $\chi^2$ increases rapidly. We obtained $\chi^2 =
135/69~{\rm DOF}$, when calculated at $\sstt = 0$, $\dms = 0$
(i.e. assuming no oscillations).

For the test of $\numu \leftrightarrow \nue$ oscillations,
we obtained a relatively poor fit; $\chi^2_{min} = 87.8 / 67
{\rm ~DOF}$, at $(\sstt = 0.93, \dms = 3.2\times10^{-3} $
eV$^2)$.
The expected asymmetry of the multi-GeV $e$-like events for the best-fit 
$\numu \leftrightarrow \nue$ oscillation hypothesis, $A=0.205$, differs
from the measured asymmetry, $A = -0.036 \pm 0.067 \pm 0.02$,
by 3.4 standard deviations.
We conclude that the $\nu_{\mu} \leftrightarrow \nu_{e}$
hypothesis is not favored. 

\begin{table}
\begin{center}
\begin{tabular}{clrl}
\multicolumn{2}{c}{Monte Carlo Fit Parameters} & Best Fit & Uncertainty \\
\hline
$\alpha$ & overall normalization & $15.8\%$ &
(*)  \\
$\delta$ & 
$E_\nu$ spectral index & 
0.006 &
$\sigma_{\delta} = 0.05$ \\
$\beta_s$  &
sub-GeV $\mu/e$ ratio & 
-6.3\% &
$\sigma_{s} = 8\%$ \\
$\beta_m$ &
multi-GeV $\mu/e$ ratio & 
-11.8\% &
$\sigma_{m} = 12\%$ \\
$\rho$ &
relative norm. of PC to FC & 
-1.8\% &
$\sigma_{\rho} = 8\%$ \\
$\lambda$ &
$L/E_\nu$ &
3.1\% &
$\sigma_\lambda = 15$\% \\
$\eta_s$ &
sub-GeV up-down &
2.4\% &
$\sigma_\eta^s = 2.4$\% \\
$\eta_m$ &
multi-GeV up-down &
-0.09\% &
$\sigma_\eta^m = 2.7$\% \\
\end{tabular}
\end{center}
\caption{Summary of Monte Carlo fit parameters. Best-fit values for 
$\numu \leftrightarrow \nutau (\dms = 2.2 \times 10^{-3} $eV$^2$, 
$\sstt = 1.0)$ and estimated uncertainties are given.
\noindent 
$^{(*)}$The over-all 
normalization ($\alpha$) was estimated to have a 25\% uncertainty but 
was fitted as a free parameter.}
\label{tb:parms}
\end{table}

\begin{figure}[t]
\begin{center} 
\psfig{figure=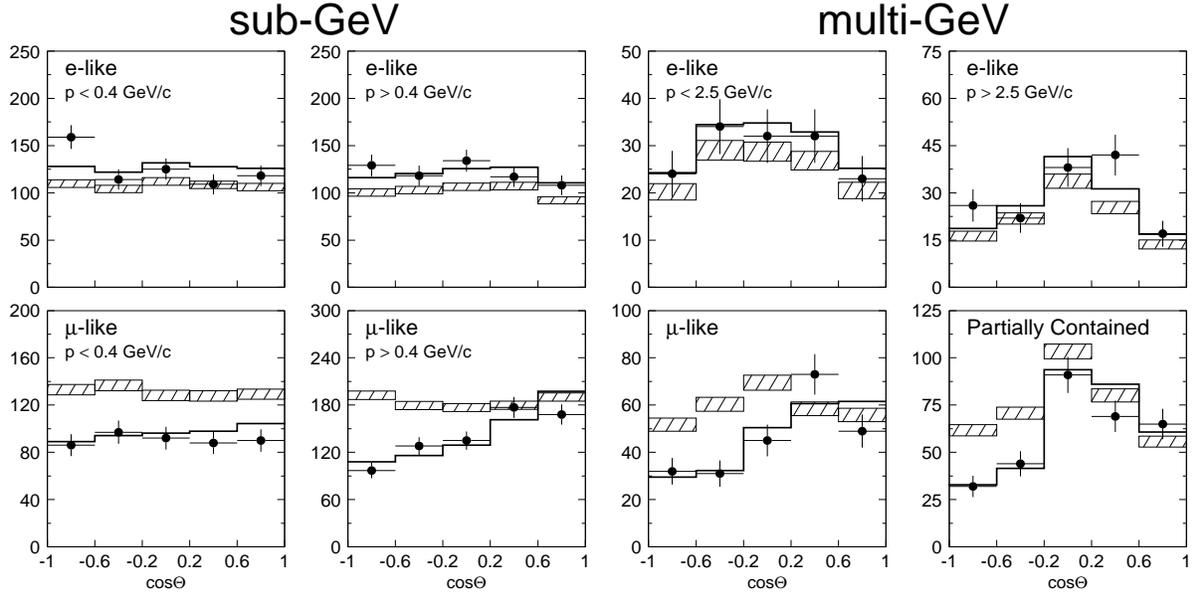,width=6.625in}
\caption{Zenith angle distributions of $\mu$-like and $e$-like events
for sub-GeV and multi-GeV data sets. Upward-going particles have $\cos
\Theta < 0$ and downward-going particles have $\cos \Theta > 0$.
Sub-GeV data are shown separately for $p < 400 $ MeV$/c$
and $p > 400 $ MeV$/c$. Multi-GeV $e$-like distributions
are shown for $p < 2.5 $ GeV$/c$  and $p > 2.5 $ 
GeV$/c$ and the multi-GeV $\mu$-like are shown
separately for FC and PC 
events. The hatched region shows the Monte Carlo expectation for no
oscillations normalized to the data live-time with statistical
errors. The bold line is the best-fit expectation for $\numu
\leftrightarrow \nutau$ oscillations with the overall flux
normalization fitted as a free parameter.}
\label{fig:angdist}
\end{center}
\end{figure}

The zenith angle distributions for the FC and PC samples are shown in
Fig.~\ref{fig:angdist}.  The data are compared to the Monte Carlo
expectation (no oscillations, hatched region) and the best-fit
expectation for $\numu \leftrightarrow \nutau$ oscillations (bold
line).

We also estimated the oscillation parameters considering 
the $R$ measurement and the zenith angle shape separately. 
The 90\% confidence level allowed regions for each case overlapped at 
$1\times10^{-3} < \dms < 4\times10^{-3}$ eV$^2$ for $\sstt = 1$. 

\begin{figure}[t]
\begin{center}
\psfig{figure=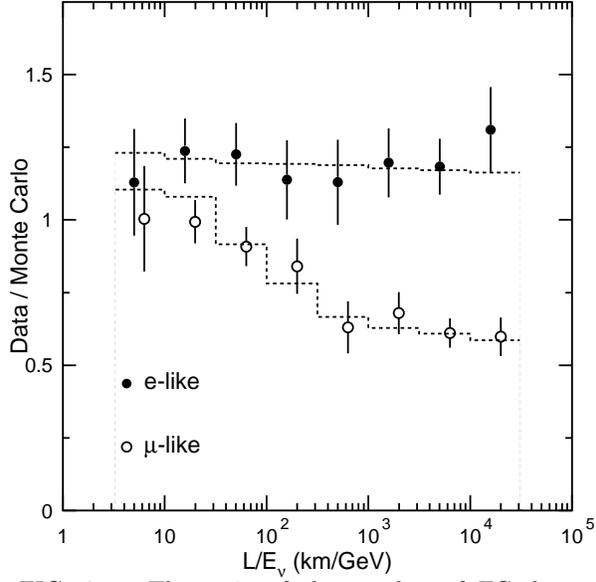,width=3.375in} 
\caption{
The ratio of the number of FC data events to FC Monte Carlo events versus
reconstructed $L/E_\nu$. The points show the ratio of observed data to MC
expectation in the absence of oscillations. The dashed lines show the
expected shape for $\numu \leftrightarrow \nutau$ at
$\dms=2.2\times10^{-3} $eV$^2$ and $\sstt=1$.
The slight $L/E_\nu$ dependence for $e$-like events is
due to contamination (2-7\%) of $\nu_\mu$ CC interactions.
}
\label{fig:loe}
\end{center}
\end{figure}

As a cross-check of the above analyses, we have reconstructed the best
estimate of the ratio $L/E_\nu$ for each event. The neutrino energy is
estimated by applying a correction to the final state lepton momentum.
Typically, final state leptons with $p\sim 100$ MeV/$c$ carry $65\%$
of the incoming neutrino energy increasing to $\sim$85\% at $p = 1$
GeV/$c$.  The neutrino flight distance $L$ is estimated following
Ref.~\cite{prodhgt} using the estimated neutrino energy and the
reconstructed lepton direction and flavor. 
Figure~\ref{fig:loe} shows the ratio of FC data to Monte
Carlo for $e$-like and $\mu$-like events with $p > 400$ MeV/$c$
as a function of $L/E_\nu$, compared to the expectation for $\numu
\leftrightarrow \nutau$ oscillations with our best-fit parameters. The
$e$-like data show no significant variation in $L/E_\nu$, while the
$\mu$-like events show a significant deficit at large $L/E_\nu$. At
large $L/E_\nu$, the $\nu_\mu$ have presumably undergone numerous
oscillations and have averaged out to roughly half the initial rate.


The asymmetry $A$ of the $e$-like events in the present data is
consistent with expectations without neutrino oscillations and
two-flavor $\nue \leftrightarrow \numu$ oscillations are not favored. 
This is in agreement with recent results from the
CHOOZ experiment\cite{chooz}. The LSND experiment has reported the
appearance of $\nue$ in a beam of $\numu$ produced by stopped
pions\cite{lsnd97}. The LSND results do not contradict the present results 
if they are observing small mixing angles. With the
best-fit parameters for $\numu \leftrightarrow \nutau$ oscillations,
we expect a
total of only 15-20 events from $\nutau$ charged-current interactions
in the data sample. Using the current sample, oscillations between
$\numu$ and $\nutau$ are indistinguishable from oscillations between
$\numu$ and a non-interacting ``sterile'' neutrino.

Figure~\ref{fig:allowed} shows the \SuperK{} results overlaid with
the allowed region obtained by the Kamiokande
experiment\cite{kam_zenith}. The \SuperK{} region favors lower values
of $\dms$ than allowed by the Kamiokande experiment; however the 90\%
contours from both experiments have a region of overlap. 
Preliminary studies of upward-going stopping 
and through-going muons in Super-Kamiokande\cite{sk_nu98} give allowed
regions consistent with the FC and PC event analysis reported in this
paper.

Both the zenith angle distribution of $\mu$-like events and the value
of $R$ observed in this experiment significantly differ from the best
predictions in the absence of neutrino oscillations. While
uncertainties in the flux prediction, cross sections, and experimental
biases are ruled out as explanations of the observations, the present
data are in good agreement with two-flavor $\numu
\leftrightarrow \nutau$ oscillations with $\sstt > 0.82$ and
$5\times10^{-4} < \dms < 6\times10^{-3}$ eV$^2$ at 90\% confidence
level. We conclude that the present data give 
evidence for neutrino oscillations.


We gratefully acknowledge the cooperation of the Kamioka Mining and
Smelting Company. The \SuperK{} experiment was built and has been
operated with funding from the Japanese Ministry of Education,
Science, Sports and Culture, and the United States Department of
Energy.

\onecolumn
\twocolumn

\end{document}